\documentclass[article,preprint,showpacs,amsfonts]{revtex4-1}
\usepackage{indentfirst}
\usepackage{amsmath,amssymb}
\def\ra{\rangle}
\def\la{\langle}

\begin{document}

\title{An alternative framework for quantifying coherence of quantum channels}
\smallskip
\author{Shi-Yun Kong$^1$}
\author{Ya-Juan Wu$^2$}
\author{Qiao-Qiao Lv$^1$}
\author{Zhi-Xi Wang$^1$}
\author{Shao-Ming Fei$^{1,3}$}
\affiliation{
{\footnotesize $^1$School of Mathematical Sciences, Capital Normal University, Beijing 100048, China}\\
{\footnotesize $^2$School of Mathematics, Zhengzhou  University of Aeronautics, Zhengzhou 450046, China}\\
{\footnotesize $^3$Max-Planck-Institute for Mathematics in the Sciences, 04103 Leipzig, Germany}
}

\begin{abstract}
We present an alternative framework for quantifying the coherence of quantum channels, which contains three conditions: the faithfulness, nonincreasing under sets of all the incoherent superchannels and the additivity. Based on the characteristics of the coherence of quantum channels and the additivity of independent channels, our framework provides an easier way to certify whether a function is a bona fide coherence measure of quantum channels. Detailed example is given to illustrate the effectiveness and advantages of our framework.
\end{abstract}

\maketitle

\section{Introduction}
As one of the characteristic features that marks the departure of quantum mechanics from the classical realm, the quantum coherence plays a central role in quantum optics \cite{glauber,scully,gyongyosi}, thermodynamics \cite{brandao,gour,narasimhachar,aberg2,lostaglio1,lostaglio2,gentaro,mayuhan,xushengzhi}, nanoscale physics \cite{plenio,rebentrost,licheming,huelga} and quantum measurements \cite{napoli,mateng,longguilu}. Quantum coherence is also the origin of many quantum phenomena such as entanglement and multiparticle interference. Recently the coherence of quantum states, concerning the quantifications \cite{baumgratz,girolami}, interconversion \cite{bromley} and applications \cite{luo,aberg,monras}, has been extensively investigated.

Similar to the resource theory of quantum entanglement, Baumgratz, Cramer and Plenio presented a rigorous framework (BCP framework) for quantifying the coherence of quantum states and introduced several measures of coherence including the relative entropy of coherence, the $l_1-$norm of coherence and fidelity \cite{baumgratz}. The BCP framework is widely used in quantifying coherence. Inspired by the BCP framework, Yu $et.~al.$ put forward another equivalent framework \cite{yuxiaodong}, which can be more conveniently used in some cases and is applicable to various physical contexts.

The quantum state transfer depends on quantum channels. The coherence of quantum channels characterizes the ability to optimize the coherence of all output states of the channels \cite{xueyuanhu,bendana,korzekawa,theurer,chandan}. Similar to the resource theory of coherence for quantum states, the resource theory of coherence for quantum channels has attracted much attention. With respect to the BCP framework for coherence of quantum states, Xu established a framework for quantifying the coherence of quantum channels \cite{xujianwei}.

In this paper, similar to the alternative framework for quantum states given in \cite{yuxiaodong} which simplifies the BCP framework in particular cases, we establish a new framework for quantifying the coherence of quantum channels, which improves the applications of the previous framework given in \cite{xujianwei}. Detailed examples are presented to illustrate the advantages of our framework.

\maketitle

\section{An alternative framework for coherence of quantum channels}

Let $H_A$ and $H_B$ be Hilbert spaces with dimensions ${\rm dim} H_A=|A|$ and ${\rm dim} H_B=|B|$, and $\{|j\ra\}_j,~\{|k\ra\}_k$ and $\{|\alpha\ra\}_\alpha,~\{|\beta\ra\}_\beta$ be the fixed orthonormal bases of $H_A$ and $H_B$, respectively. Denote $\mathcal{D}_A$ ($\mathcal{D}_B$) the set of all density operators on $H_A$ ($H_B$). Let $\mathcal{C}_{AB}$ be the set of all channels from $\mathcal{D}_A$ to $\mathcal{D}_B$.

A quantum channel $\phi\in \mathcal{C}_{AB}$ is a linear completely positive and trace preserving (CPTP) map with Kraus operators $\{K_n\}_n$ satisfying $\sum_n K_n^\dagger K_n=I$ such that $\phi(\rho)=\sum_n K_n\rho K_n^\dagger$. The corresponding Choi matrix with respect to the channel $\phi\in\mathcal{C}_{AB}$ has the following form,
\begin{equation}\label{W1}
J_{\phi}=\sum_{jk}|j\ra\la k|\otimes\phi(|j\ra\la k|)=\sum_{jk,\alpha\beta}\phi_{jk\alpha\beta}|j\ra\la k|\otimes|\alpha\ra\la\beta|,
\end{equation}
where $\phi_{jk\alpha\beta}=\la\alpha|\phi(|j\ra\la k|)|\beta\ra$ are complex numbers and $\sum_{\alpha}\phi_{jk,\alpha\alpha}=\delta_{jk}$.
If $C$ is a coherence measure of quantum states, then $C(\phi)=C(\frac{J_\phi}{|A|})$ is a corresponding coherence measure for quantum channels \cite{xujianwei}, which quantifies the coherence of the channel by the coherence of the state $\frac{J_\phi}{|A|}$. The Choi matrix of a channel gives a relation between the coherence measures for quantum channels and for that quantum states.

Let $\mathcal{I}$ be the set of all incoherent states whose density matrices are diagonal in the given basis. An incoherent state $\rho\in H_A$ satisfies $\Delta_A(\rho)=\rho$ with $\Delta_A(\rho)=\sum_j\la j|\rho|j\ra|j\ra\la j|$ being a completely  dephasing channel. $\phi\in\mathcal{C}_{AB}$ is called an incoherent quantum channel if $\Upsilon(\phi)=\phi$, where $\Upsilon(\phi)=\Delta_B \phi \Delta_A$, $\Delta_A$ and $\Delta_B$ are resource destroying maps\cite{liuziwen}. We denote $\mathcal{IC}$ the set of all the incoherent channels.
Let $\mathcal{SC}_{ABA'B'}$ be the superchannels that are linear maps from $\mathcal{C}_{AB}$ to $\mathcal{C}_{A'B'}$, we define the Choi matrix of the superchannel $\Theta\in\mathcal{SC}_{ABA'B'}$ as $J_\Theta=\sum_{jk\alpha\beta}|j\alpha\ra\la k\beta|\otimes\Theta(|j\alpha\ra\la k\beta|)$. Any superchannel $\Theta$ is an incoherent superchannel ($\mathcal{ISC}$) if there exists an expression of Kraus operators $\Theta=\{M_m\}_m$ such that for each $m$, $M_m=\sum_{j\alpha}M_{mj\alpha}|f(j\alpha)\ra\la j\alpha|$ with $f(j\alpha)=f(j,\alpha)\in \{(j',\alpha')|_{j'=1}^{|A'|},_{\alpha'=1}^{|B'|}\}$.

In \cite{xujianwei}, the author presented a framework for quantifying the coherence of quantum channels. A proper measure $C$ of the coherence for quantum channels must satisfy
the following conditions:
\begin{enumerate}
\item[$\mathrm{(B1)}$] For any quantum channel $C(\phi)\geqslant0$, $C(\phi)=0$ if and only if $\phi\in\mathcal{IC}_{AB}$;
\item[$\mathrm{(B2)}$] $C(\phi)\geqslant C[\Theta(\phi)]$ for any incoherent superchannel $\Theta$;
\item[$\mathrm{(B3)}$] $C(\phi)\geqslant\sum_m p_mC(\phi_m)$ for any incoherent superchannel $\Theta$, with $\{M_m\}_m$ an incoherent Kraus operator of $\Theta$, $p_m=\frac{\mathrm{tr}(M_m J_\phi M_m^\dagger)}{|A'|}$, and $J_{\phi_m}=|A'|\frac{M_m J_\phi M_m^\dagger}{\mathrm{tr}(M_m J_\phi M_m^\dagger)}$;
\item[$\mathrm{(B4)}$] $C(\sum_m p_m \phi_m)\leqslant\sum_m p_m C(\phi_m)$ for any set of quantum channels $\{\phi_m\}$ and any probability distribution $\{p_m\}_m$.
\end{enumerate}
The items (B1)-(B4) give necessary conditions for a bona fide measure of coherence for quantum channels, from which the quantification of the coherence for quantum channels has been further
investigated.

Nevertheless, similar to the case of coherence measures for quantum states, the last two conditions (B3) and (B4) are rather difficult to be verified for a given measure of coherence
for quantum channels. In the following we present an alternative framework consisting of three conditions, which is equivalent to the above framework but can be easily applied.
A function $C$ is a well defined coherence measure of quantum channels if it satisfies the following three conditions:
 \begin{enumerate}
\item[$\mathrm{(C1)}$] $\mathit{Faithfulness}$. $C(\phi)\geqslant0$ for any $\phi\in \mathcal{C}_{AB}$, and $C(\phi)=0$ if and only if $\phi\in\mathcal{IC}_{AB}$;
\item[$\mathrm{(C2)}$] $\mathit{Nonincreasing\ under\ ISCs}.$ $C(\phi)\geqslant C[\Theta(\phi)]$ for any $\Theta\in\mathcal{ISC}_{ABA'B'}$;
\item[$\mathrm{(C3)}$] $\mathit{Additivity}$. $C(\Phi)=p_1C(\phi_1)+p_2C(\phi_2)$ for $p_1+p_2=1$, $\phi_1\in \mathcal{C}_{AB_1}$ and $\phi_2\in \mathcal{C}_{AB_2}$, where $\Phi(|j\ra\la k|)=p_1\phi_1(|j\ra\la k|)\oplus p_2\phi_2(|j\ra\la k|)$, $\Phi\in \mathcal{C}_{AB}$,\ and $|B|=|B_1|+|B_2|$.
\end{enumerate}

In the following, we prove that the framework given by (B1)-(B4) is equivalent to the one given by (C1)-(C3). We first prove that (B1)-(B4) give rise to (C1)-(C3), namely, (B3)(B4) give rise to (C3) since (C1) and (C2) are the same as (B1) and (B2).
Consider a CPTP map $\Theta_1\in\mathcal{ISC}_{ABAB}$, $\Theta_1(\cdot)=Q_1\cdot Q_1^\dagger+Q_2\cdot Q_2^\dagger$, where
\begin{equation}
\begin{aligned}
Q_1=&|0\ra\la0|+\cdots+||B_1|-1\ra\la |B_1|-1|+||B_1|+|B_2|\ra\la |B_1|+|B_2||+\cdots+|2|B_1|+|B_2|-1\ra\\
&\la 2|B_1|+|B_2|-1|+\cdots+|(|A|-1)(|B_1|+|B_2|)\ra\la(|A|-1)(|B_1|+|B_2|)|+\cdots\\
&+|(|A|-1)(|B_1|+|B_2|)+|B_1|-1\ra\la(|A|-1)(|B_1|+|B_2|)+|B_1|-1|
\end{aligned}
\end{equation}
and
\begin{equation}
\begin{aligned}
Q_2=&||B_1|\ra\la |B_1||+\cdots+||B_1|+|B_2|-1\ra\la |B_1|+|B_2|-1|+|2|B_1|+|B_2|\ra\la2|B_1|+|B_2||+\cdots\\
&+|2(|B_1|+|B_2|)-1\ra\la2(|B_1|+|B_2|)-1|+\cdots+||A||B_1|+(|A|-1)|B_2|\ra\\
&\la |A||B_1|+(|A|-1)|B_2||+\cdots+||A|(|B_1|+|B_2|)-1\ra\la |A|(|B_1|+|B_2|)-1|.
\end{aligned}
\end{equation}
Note that $Q_1$ and $Q_2$ are just projectors onto $\mathcal{C}_{AB}$. Obviously, one sees that $Q_i\mathcal{IC}Q_i^\dagger\subset\mathcal{IC}$. The Choi matrix of $\Phi$ in (C3) is given by \begin{equation}\label{W3}
J_\Phi=\sum_{j,k}|j\ra\la k|\otimes[p_1\phi_1(|j\ra\la k|)\oplus p_2\phi_2(|j\ra\la k|)].
\end{equation}
Then
\begin{equation}
\Theta_1(\Phi)=Q_1J_\Phi Q_1^\dagger+Q_2J_\Phi Q_2^\dagger=p_1J_{\tilde{\phi}_1}+p_2J_{\tilde{\phi}_2},
\end{equation}
where $p_1=\frac{\mathrm{tr}(Q_1J_\Phi Q_1^\dagger)}{|A|}$, $p_2=\frac{\mathrm{tr}(Q_2J_\Phi Q_2^\dagger)}{|A|}$ and $J_{\tilde{\phi}_1}=|A|\frac{Q_1J_\Phi Q_1^\dagger}{\mathrm{tr}(Q_1J_\Phi Q_1^\dagger)}$, $J_{\tilde{\phi}_2}=|A|\frac{Q_2J_\Phi Q_2^\dagger}{\mathrm{tr}(Q_2J_\Phi Q_2^\dagger)}$. From (B2) and (B3) we have
\begin{equation}\label{W4}
C(\Phi)\geqslant p_1C(\tilde{\phi_1})+p_2C(\tilde{\phi_2}),
\end{equation}
where $\tilde{\phi}_1,\,\tilde{\phi}_2\in\mathcal{C}_{AB}$, $\tilde{\phi}_1(|j\ra\la k|)=\phi_1(|j\ra\la k|)\oplus{\bf0}(|j\ra\la k|)$ and $\tilde{\phi}_2(|j\ra\la k|)={\bf0}(|j\ra\la k|)\oplus\phi_2(|j\ra\la k|)$ with ${\bf0}$ a zero map.
From (B4) we have
\begin{equation}\label{W5}
C(\Phi)\leqslant p_1C(\tilde{\phi_1})+p_2C(\tilde{\phi_2}).
\end{equation}
Combining \eqref{W4} with \eqref{W5}, we get
\begin{equation}\label{W6}
C(\Phi)=p_1C(\tilde{\phi_1})+p_2C(\tilde{\phi_2}).
\end{equation}

To obtain (C3), we need to certify $C(\tilde{\phi_1})=C(\phi_1)$ further.
Under an incoherent superchannel, any $\phi$ and $\overline{\Theta}(\phi)=\sum P_nJ_\phi P_n^\dagger$ can be transformed into each other, where the Kraus operators $\{P_n\}_n$ are permutation matrices. By (B2) we have $C(\overline{\Theta}(\phi))=C(\phi)$. Based on this fact, we define an incoherent superchannel $\overline{\Theta}_1\in\mathcal{ISC}_{ABAB}$ with Kraus operators $\{P_{nl}^{(1)}\}_{nl}$ as the permutation matrices,
\begin{equation}
P_{nl}^{(1)}(i,j)=\left\{
\begin{aligned}
&1,~{\rm if}~(i, j)=(i_{nl},  j_{nl}) ~{\rm or} ~(i, j)=(j_{nl},  i_{nl}),\\
&1,~{\rm if}~i=j,\\
&0,~\rm{other wise},
\end{aligned}\right.
\end{equation}
where $i_{nl}=(n-1)(|B_1|+|B_2|)+l$, $j_{nl}=(n-1)|B_1|+l$, $i,j=1,\cdots,|A||B|$, $n=1,\cdots,|A|$ and $l=1,\cdots,|B_1|$.
Then, \begin{equation}\overline{\Theta}_1(\tilde{\phi}_1)=\sum_{n,l}P_{nl}^{(1)}J_{\tilde{\phi}_1}P_{nl}^{(1)\dagger}=J_{\phi_1}\oplus O_{AB_2},\end{equation}
where $O_{AB_2}$ is a $|A||B_2|\times|A||B_2|$ null matrix. It is easily seen that
\begin{equation}\label{W7}
C(\tilde{\phi}_1)=C(\overline{\Theta}_1(\tilde{\phi}_1)).
\end{equation}

Next, we need to prove $C(\overline{\Theta}_1(\tilde{\phi}_1))=C(\phi_1)$. For this, we define two incoherent superchannels: $\Theta_2\in\mathcal{ISC}_{AB_1AB}$ with Kraus operator $M_0$ satisfying $\la j|M_0|k\ra=\delta_{jk}$ and $\Theta_3\in\mathcal{ISC}_{ABAB_1}$ with Kraus operators $\{M_n\}_{n=0}^{\lceil\frac{|B_2|}{|B_1|}\rceil}$ satisfying $\la j|M_n|k\ra=\delta_{j,k-n|B_1|}$. Then we get, \begin{equation}~\Theta_2(\phi_1)=M_0J_{\phi_1}M_0^\dagger=J_{\phi_1}\oplus O_{AB_2}\end{equation}
and
\begin{equation}
\Theta_3[(\overline{\Theta}_1(\tilde{\phi}_1)]
=\sum_{n=0}^{\lceil\frac{|B_2|}{|B_1|}\rceil}M_n(J_{\phi_1}\oplus O_{AB_2})M_n^\dagger=J_{\phi_1}.
\end{equation}
From (B2) we obtain
\begin{equation}\label{6jia}
C(\overline{\Theta}_1(\tilde{\phi}_1))=C(\phi_1).
\end{equation}
Combining \eqref{6jia} with \eqref{W6} and \eqref{W7}, we get the condition (C3).

We have shown that any $C$ satisfying (B1)-(B4) also satisfies (C1)-(C3). Next, we prove that any $C$ satisfying (C1)-(C3) must satisfy (B3) and (B4).
First, we prove (B3), i.e., $C$ is convex. Define $\Phi_1\in\mathcal{C}_{AB'}$ as,
\begin{equation}\Phi_1(|j\ra\la k|)=\phi(|j\ra\la k|)\oplus {\bf 0}(|j\ra\la k|)\oplus\cdots\oplus {\bf 0}(|j\ra\la k|),\end{equation}
where $\phi\in\mathcal{C}_{AB},\ H_{B'}=\underbrace{H_B\otimes H_B\otimes\cdots\otimes H_B}_{M}.$
From (C3), we have
\begin{equation}\label{W8}
C(\Phi_1)=C(\phi).
\end{equation}
Consider $\overline{\Theta}_2\in\mathcal{ISC}_{AB'AB'}$, with its Kraus operators $\{P_n^{(2)}\}_n$ being the permutation matrices, such that
$\overline{\Theta}_2(\Phi_1)=\sum_n P_n^{(2)}J_{\Phi_1}P_n^{(2)\dagger}=J_\phi\oplus \underbrace{O_{AB}\oplus\cdots\oplus O_{AB}}_{M-1}$. Apply an incoherent superchannel $\Theta_4\in \mathcal{SC}_{ABA_1B'_{1}}$ with Kraus operators $\{U_m\otimes M_m\}_m$ such that
\begin{equation}
\Theta_4[\overline{\Theta}_2(\Phi_1)]=\sum_{m=0}^{M-1}(U_m\otimes M_m)(\sum_nP_n^{(2)}J_{\Phi_1}P_n^{(2)\dagger})(U_m\otimes M_m)^\dagger,
\end{equation}
where $U_m=\sum_{k=0}^{M-1}|(k+m)~\mathrm{mod}~M\ra\la k|,~H_{B'_1}=\underbrace{H_{B_1}\otimes H_{B_1}\otimes\cdots\otimes H_{B_1}}_{M}$, and $\{M_m\}$ are incoherent Kraus operators of the superchannel in $\mathcal{ISC}_{ABA_1B_1}$. One can easily see that $(U_m\otimes M_m)\mathcal{IC}(U_m\otimes M_m)^\dagger \subset\mathcal{IC}$,
\begin{equation}
\Theta_4[\overline{\Theta}_2(\Phi_1)]=\sum_{m=0}^{M-1}p_m |m\ra\la m|\otimes J_{\phi_m},
\end{equation}
where $p_m=\frac{\rm{tr}(M_mJ_\phi M_m^\dagger)}{|A_1|}$ and $J_{\phi_m}=|A_1|\frac{M_mJ_\phi M_m^\dagger}{\rm{tr}(M_mJ_\phi M_m^\dagger)}$.

Similarly, there exists $\overline{\Theta}_3\in\mathcal{ISC}_{A_1B_1'A_1B_1'}$, with its Kraus operators $\{P_n^{(3)}\}_n$ being the permutation matrices, such that $\overline{\Theta}_3[\Theta_4(\overline{\Theta}_2(\Phi_1))]=\sum P_n^{(3)}(\sum_{m=0}^{M-1}p_m |m\ra\la m|\otimes J_{\phi_m})P_n^{(3)\dagger}=\sum_{i,j=0}^{M-1}|i\ra\la j|\otimes[p_0\phi_0(|i\ra\la j|)\oplus\cdots\oplus p_{M-1}\phi_{M-1}(|i\ra\la j|)]$. $\overline{\Theta}_3[\Theta_4(\overline{\Theta}_2(\Phi_1))]$ matches to a channel $\Phi_2\in C_{A_1{B_1}'},~\Phi_2(|i\ra\la j|)=p_0\phi_0(|i\ra\la j|)\oplus\cdots\oplus p_{M-1}\phi_{M-1}(|i\ra\la j|)$.
\\Following (C3), we have
\begin{equation}\label{W9}
C(\Phi_2)=\sum_{m=0}^{M-1}p_m C(\phi_m).
\end{equation}
By (C2), \eqref{W8} and \eqref{W9} we can have\begin{equation}C(\phi)\geqslant\sum_{m=0}^{M-1}p_m C(\phi_m),
\end{equation}
which proves that (B3) holds.

We now prove (B4). We first define an initial channel $\Phi_3\in C_{AB'}$ satisfying \begin{equation}\Phi_3(|j\ra\la k|)=\oplus_{m=0}^{M-1}p_m\phi_m(|j\ra\la k|),\end{equation}
where $\phi_m\in\mathcal{C}_{AB},~\{p_m\}$ are the probability distribution of $\{\phi_m\}$ and $\sum_{m=0}^{M-1}p_m=1$.
According to (C3), one has
\begin{equation}\label{W10}
C(\Phi_3)=\sum_{m=0}^{M-1}p_m C(\phi_m).
\end{equation}
Apply $\overline{\Theta}_4\in\mathcal{ISC}_{AB'AB'}$, with its Kraus operators being the permutation matrices $\{P_n^{(4)}\}$, such that $\overline{\Theta}_4(\Phi_3)=\sum_n P_n^{(4)} J_{\Phi_3} P_n^{(4)\dagger}=\oplus_{m=0}^{M-1}p_m J_{\phi_m}$.
Let $\Theta_5\in\mathcal{ISC}_{AB'AB'}$ be an incoherent super channel such that
\begin{equation}\begin{aligned}
\Theta_5\overline{\Theta}_4(\Phi_3)&=\sum_{m=0}^{M-1}(|0\ra\la m|\otimes I)(\sum_n P_n^{(4)} J_{\Phi_3} P_n^{(4)\dagger})(|0\ra\la m|\otimes I)^\dagger\\
&=\sum_{m=0}^{M-1}p_mJ_{\phi_m}\oplus O_{A_1B_1}\oplus\cdots\oplus O_{A_1B_1}.\end{aligned}
\end{equation}
Apply $\overline{\Theta}_5\in\mathcal{ISC}_{AB'AB'}$, with Kraus operators $\{P_n^{(5)}\}$ as permutation matrices, such that
\begin{equation}\begin{aligned}
\overline{\Theta}_5[\Theta_5\overline{\Theta}_4(\Phi_3)]&=\sum_n P_n^{(5)}(\sum_{m=0}^{M-1}p_mJ_{\phi_m}\oplus O_{A_1B_1}\oplus\cdots\oplus O_{A_1B_1})P_n^{(5)\dagger}\\
&=\sum_{j,k=0}^{|A|-1}|j\ra\la k|\otimes[\sum_{m=0}^{M-1}p_m\phi_m(|j\ra\la k|)\oplus{\bf 0}(|j\ra\la k|)\oplus\cdots\oplus{\bf0}(|j\ra\la k|)].
\end{aligned}\end{equation}
Thus, $\overline{\Theta}_5[\Theta_5\overline{\Theta}_4(\Phi_3)]$ corresponds to $\Phi_4\in\mathcal{C}_{AB'}$ with $\Phi_4(|j\ra\la k|)=\sum_{m=0}^{M-1}p_m\phi_m(|j\ra\la k|)\oplus{\bf0}(|j\ra\la k|)\oplus\cdots\oplus{\bf0}(|j\ra\la k|)$. From (C3) we have
\begin{equation}\label{W11}
C(\Phi_4)=C(\sum_{m=0}^{M-1}p_m\phi_m).
\end{equation}
Combining (C2) with \eqref{W10} and \eqref{W11}, we get \begin{equation}
\sum_{m=0}^{M-1}p_mC(\phi_m)\geqslant C(\sum_{m=0}^{M-1}p_m\phi_m),
\end{equation}
namely, (B4) holds.

We usually get coherence measures for quantum channels from corresponding coherence measures for quantum states. For instance, the $l_1-$norm of coherence $C_{l_1}(\rho)=\sum_{i\ne j}|\rho_{i,j}|$\cite{baumgratz} and the relative entropy of coherence $C_{\rm rel.}(\rho)=S(\rho_{\rm diag})-S(\rho),$ where $S$ is the von Neumann entropy and $\rho_{\rm diag}$ denotes the state obtained from $\rho$ by deleting all off-diagonal elements\cite{baumgratz}, are coherence measures for quantum states, on this basis, $C_{l_1}(\phi)=\sum_{i\ne j}|\frac{J_\phi}{|A|}|$ and $C_{\rm rel.}(\phi)=S(\phi_{\rm diag})-S(\phi)=S(\frac{J_{\phi_{\rm diag}}}{|A|})-S(\frac{J_\phi}{|A|})$ both are coherence measures for quantum channels\cite{xujianwei}.

The above proof shows that our new framework is equivalent to the framework given by (B1)-(B4) for quantum channels. In determining whether a function $C$ can be used as a coherence measure for channels, in some cases, it is not easy to verify whether $C$ satisfies (B3). The condition (C3) in our framework provides a new way to solve the problem. We give an example to show the efficiency of our framework.

{\bf Example}
The trace distance measure of coherence defined by $C_{\rm tr}(\rho):=\min_{\delta \in  \cal{I}} \|\rho-\delta\|_{\rm tr}=\min_{\delta \in  \cal{I}}{\rm tr}|\rho-\delta|$ is not a well defined coherence measure for quantum states \cite{yuxiaodong}. Let us check whether $C_{\rm tr}(\phi)=C_{\rm tr}(\frac{J_\phi}{|A|})$ is a bona fide coherence measure for quantum channels or not.
Here we define
\begin{equation}
C_{\rm tr}(\phi):=\min_{{\tilde{\phi}}\in \cal{IC}}\|\phi-\tilde{\phi}\|_{\rm tr},
\end{equation}
where $\|\phi-\tilde{\phi}\|_{\rm tr}=\|\frac{J_\phi}{|A|}-\frac{J_{\tilde{\phi}}}{|A|}\|_{\rm tr}={\rm tr}|\frac{J_\phi}{|A|}-\frac{J_{\tilde{\phi}}}{|A|}|$ is the trace norm between $\phi$ and $\tilde{\phi}$ with $\phi,\,\tilde{\phi}\in \mathcal{C}_{AB}$.
We need to verify that $C_{\rm tr}(\phi)$ satisfies either (B1)-(B4) or (C1)-(C3). It has been already proved in previous works \cite{baumgratz,bromley} that $C_{\rm tr}$ satisfies (B1), (B2) and (B4). However, the verification of (B3) is rather difficult. The inequality can only be fulfilled for qubit and $X$ quantum states \cite{shaolianhe,Rana}.

We use condition (C3) to verify the validity of $C_{\rm tr}(\phi)$.
For the isometry channel $\phi_{\rm max}\in \mathcal{C}_{AB}$ \cite{xujianwei},
\begin{equation}\phi_{\rm max}(|j\ra\la k|)=\frac{1}{|B|}\sum_{\alpha, \beta=0}^{|B|-1}e^{i(\theta_{j\alpha}-\theta_{k\beta})}|\alpha\ra\la\beta|,\end{equation}
we have
\begin{equation}C_{\rm tr}(\phi_{\rm max})=\min_{\tilde{\phi}\in\mathcal{IC}}\|\frac{J_{\phi_{\rm max}}}{|A|}-\frac{J_{\tilde{\phi}}}{|A|}\|_{\rm tr},\end{equation}
where
\begin{equation}\frac{J_{\phi_{\rm max}}}{|A|}=\frac{1}{|A||B|}\sum_{j,k=0}^{|A|-1}|j\ra\la k|\otimes(\sum_{\alpha,~\beta=0}^{|B|-1}e^{i(\theta_{j\alpha}-\theta_{k\beta})}|\alpha\ra\la\beta|)=|\psi\ra\la\psi|,
\end{equation}
with $|\psi\ra=\frac{1}{\sqrt{|A||B|}}\sum_{j=0}^{|A|-1}\sum_{\alpha=0}^{|B|-1}e^{i\theta_{j\alpha}}|j\alpha\ra$.

Set $U_n=\sum_{k=0}^{|A||B|-1}e^{i(\theta_{(k+n)~\rm mod~|A||B|}-\theta_{k})}|(k+n)~{\rm mod}~|A||B|\ra\la k|$. Then we have $U_n|\psi\ra=|\psi\ra$.
Since $\|A\|_{\rm tr}+\|B\|_{\rm tr}\geqslant\|A+B\|_{\rm tr}$ and $\|U_n|\psi\ra\|_{\rm tr}=~\||\psi\ra\|_{\rm tr}$ for the unitary operation $U_n$, we obtain
\begin{equation}
\begin{array}{rcl}
\|\frac{J_{\phi_{\rm max}}}{|A|}-\frac{J_{\tilde{\phi}}}{|A|}\|_{\rm tr}&=&\frac{1}{|A||B|}\sum_{n=0}^{|A||B|-1}\|U_n(\frac{J_{\phi_{\rm max}}}{|A|}-\frac{J_{\tilde{\phi}}}{|A|}){U_n}^\dagger\|_{\rm tr}\\[1mm]
&\geqslant&\frac{1}{|A||B|}\|\sum_{n=0}^{|A||B|-1}(U_n(\frac{J_{\phi_{\rm max}}}{|A|}-\frac{J_{\tilde{\phi}}}{|A|}){U_n}^\dagger)\|_{\rm tr}.
\end{array}
\end{equation}
As
\begin{equation}
U_n\frac{J_{\phi_{\rm max}}}{|A|}{U_n}^\dagger=U_n|\psi\ra\la\psi|{U_n}^\dagger=\frac{J_{\phi_{\rm max}}}{|A|}
\end{equation}
and
\begin{equation}\sum_{n=0}^{|A||B|-1}U_n\frac{J_{\tilde{\phi}}}{|A|}{U_n}^\dagger=I_{|A||B|},
\end{equation}
we have
\begin{equation}
\|\frac{J_\phi}{|A|}-\frac{J_{\tilde{\phi}}}{|A|}\|_{\rm tr}\geqslant\|\frac{J_\phi}{|A|}-\frac{1}{|A||B|}I_{|A||B|}\|_{\rm tr}.
\end{equation}
Therefore,
\begin{equation}\label{12}
\min_{\tilde{\phi}\in\mathcal{IC}}\|\frac{J_{\phi_{\rm max}}}{|A|}-\frac{J_{\tilde{\phi}}}{|A|}\|_{\rm tr}=\|\frac{J_{\phi_{\rm max}}}{|A|}-\frac{1}{|A||B|}I_{|A||B|}\|_{\rm tr}=\frac{2(|A||B|-1)}{|A||B|}.
\end{equation}

Next, we consider a specific channel $\phi\in \mathcal{C}_{AB}$,
\begin{equation}
\phi(|j\ra\la k|)=\frac{1}{2}\phi_1(|j\ra\la k|)\oplus\frac{1}{2}\phi_2(|j\ra\la k|),
\end{equation}
where
\begin{equation}
\phi_1(|j\ra)=\frac{1}{\sqrt{2}}\sum_{\alpha=0}^{1}e^{i\theta_{j\alpha}}|\alpha\ra,
~~~\phi_2(|j\ra)=\frac{1}{\sqrt{3}}\sum_{\beta=0}^{2}e^{i\theta_{j\beta}}|\beta\ra,
\end{equation}
with $\phi_1\in\mathcal{C}_{AB_1}$ and $\phi_2\in\mathcal{C}_{AB_2}$ the isometry channels, $|A|=2,~|B_1|=2,~|B_2|=3$ and $|B|=5$.
In particular, we take the incoherent channel $\phi_0\in C_{AB},~\phi_0(|j\ra\la k|)=\frac{1}{4}\delta_{jk}(|j\ra\la k|)\oplus \mathbf{0}(|j\ra\la k|)$. Then
\begin{equation}
C_{\rm tr}(\phi)=\displaystyle\min_{\tilde{\phi}\in\mathcal{IC}}\|\frac{J_\phi}{2}-\frac{J_{\tilde{\phi}}}{2}\|_{\rm tr}\leqslant\|\frac{J_\phi}{2}-\frac{J_{\phi_0}}{2}\|_{\rm tr}.
\end{equation}
From (10) we get $\frac{1}{2}C_{\rm tr}(\phi_1)+\frac{1}{2}C_{\rm tr}(\phi_2)=\frac{19}{12}$. However, $C_{\rm tr}(\phi)\leqslant\|\frac{J_\phi}{2}-\frac{J_{\phi_0}}{2}\|_{\rm tr}=1$. Obviously, $C_{\rm tr}(\phi)\ne\frac{1}{2}C_{\rm tr}(\phi_1)+\frac{1}{2}C_{\rm tr}(\phi_2)$. Therefore, the trace norm of coherence is not a well defined coherence measure of quantum channels. In other words, it also violates (B3). Here, inspired by the definition of trace norm, one may propose a similar trace norm function
${C_{\rm tr}}'(\phi)=\displaystyle\min_{\lambda\geqslant0,
~\tilde{\phi}\in\mathcal{IC}}\|\phi-\lambda\tilde{\phi}\|_{\rm tr}$, which can be shown to be a legal coherent measure for quantum channels \cite{yuxiaodong,xujianwei}.

We have studied the coherence of quantum channels based on the corresponding Choi matrices to the quantum channels. Note that $p_1J_{\phi_1}\oplus p_2J_{\phi_2}$ is not necessarily a Choi matrix for arbitrary channels $\phi_1\in \mathcal{C}_{AB_1}$ and $\phi_2\in \mathcal{C}_{AB_2}$. From (1) the Choi matrix corresponding to a channel $\phi\in \mathcal{C}_{AB}$ is a $|A||B|\times|A||B|$ positive definite matrix where each $\phi(|j\ra\la k|)$ is a $|B|\times|B|$ block matrix and ${\rm tr}(\phi(|j\ra\la j|))=1$. Assuming that there is a channel $\Phi\in \mathcal{C}_{AB}$ such that $J_{\Phi}=p_1J_{\phi_1}\oplus p_2J_{\phi_2}$ with $|B|=|B_1|+|B_2|$. With respect to the matrix $p_iJ_{\phi_i}$, each $p_i\phi_i(|j\ra\la k|)$ is a $|B_i|\times|B_i|$ block matrix and ${\rm tr}[p_i\phi_i(|j\ra\la j|)]=p_i$, $i=1,2$. One can see that the trace of the $|B|\times|B|$ block matrix on all diagonals cannot always be 1 for arbitrary probability $p_1$ and $p_2$. In other words, $p_1J_{\phi_1}\oplus p_2J_{\phi_2}$ is not necessarily a Choi matrix, as it does not satisfy the structure of the Choi matrix corresponding to the channels.

\section{Conclusions}

We have presented an alternative framework to quantify the coherence of quantum channels. Our framework and the framework given by (B1)-(B4) for quantum channels are equivalent. We have used this framework to certify the validity of the trace norm coherence measure for quantum channels. Similar to the case for the coherence measure of quantum states \cite{yuxiaodong}, our framework has the similar unique superiorities and may significantly simplify the quantification for coherence of quantum channels. Our results may highlight further investigations on the resource theory of quantum channels.

\bigskip
\noindent{\bf Acknowledgments}\, \, This work is supported by NSFC (Grant Nos. 12075159 and 12171044), Beijing Natural Science Foundation (Z190005), Academy for Multidisciplinary Studies, Capital Normal University, the Academician Innovation Platform of Hainan Province, and Shenzhen Institute for Quantum Science and Engineering, Southern University of Science and Technology (No. SIQSE202001); Academician Innovation Platform of Hainan Province.

\end{document}